\documentclass{PHYEAUTH}

\usepackage{amsmath}
\usepackage{amssymb}
\usepackage{graphicx}
\usepackage{dcolumn}
\usepackage{bm}

\def\be{\begin{eqnarray}}
\def\ee{\end{eqnarray}}
\def\ba{\begin{array}}
\def\ea{\end{array}}

\begin{document}
\begin{frontmatter}
\title{Investigating the current distribution of parallel-configured quantum point contacts under quantum Hall conditions}
\author[l1]{Deniz Eksi}
and
\author[l2,l3]{Afif Siddiki}

\address[l1]{Vacational School of Health, Yeni Yuzyil University, Istanbul, 34010, Turkey}
\address[l2]{Maltepe University, Faculty Engineering and Natural Sciences, Department of Electrics and Electronics, 34857  Istanbul,  Turkey}
\address[l3]{Ekendiz Tanay Center for Art and Science, Department of Physics, Ula, Mugla, 48650, Turkey}




\begin{abstract}
Electric-field-controlled charge transport is a key concept of modern computers, embodied namely in field effect transistors. The metallic gate voltage controls charge population, thus it is possible to define logical elements which are the key to computational processes. Here, we investigate a similar system defined by metallic gates inducing quasi-one-dimensional transport channels on a high-mobility electron system in the presence of a strong perpendicular magnetic field. Firstly, we solve the three-dimensional Poisson equation, self-consistently imposing relevant boundary conditions, and use the output as an initial condition to calculate charge density and potential distribution in the plane of a two-dimensional electron system, in the presence of an external magnetic field. Subsequently, we impose an external current and obtain the spatial distribution of the transport charges, considering various magnetic field and gate voltage strengths at sufficiently low ($<$ 10 Kelvin) temperatures. We show that magnetic field breaks the spatial symmetry of the current distribution, whereas voltage applied to metallic gates determines the scattering processes.
\end{abstract}


\end{frontmatter}


\section{Introduction}

Discovery of semiconductor-based electronics stemming from quantum mechanics revolutionized our computational abilities~\cite{Davies}. The basic idea behind this is to confine electrons in the growth direction ($z$) to a plane and control their population by an electric field applied to the metallic gates residing on the surface. These structures are known as the field-effect transistors (FETs). The best known of these semiconductor devices are the metal-oxide-silicon (MOS) heterojunctions, which are the main ingredients of our daily used computers. A similar heterostructure is the GaAs/AlGaAs junction, in which the electron mobility is much higher~\cite{Datta}, i.e., scattering due to impurities is reduced. Here, at the initial crystal growth the average electron density $n_{\rm el}$ is fixed by the number of silicon donors $n_0$ which are homogeneously distributed, and electrons are confined to a single quantum well, forming a two-dimensional electron system (2DES). In this paper, we focus on such high-mobility 2DESs, where charge transport is also controlled by surface gates.

The above described 2DESs present peculiar transport properties when they are subject to high and perpendicular magnetic fields $B$, known as quantum Hall effects~\cite{Girvin00:book}, the study of which has produced two Nobel prizes. It is observed that the longitudinal resistance vanishes at certain $B$ intervals, whereas the transverse (Hall) resistance assumes quantized values in units of conductance quanta $e^2/h$~\cite{vK80:494}. Moreover, even in the absence of an external $B$ field, gate-voltage-induced narrow transport channels also present quantized conductance behavior~\cite{Wees88:848}. Such devices are named quantum point contacts (QPCs), which are the main object of our investigation~\cite{Kristensen98:180,SiddikiMarquardt,Arslan08:125423}. These devices are claimed to be a key element in developing quantum computers, while coherence is a significant parameter in charge transport, and topologically protected information processing is required~\cite{AdyStern:quantumcomp}.

The scope of this paper is to provide a self-consistent calculation scheme which is able to describe electronic transport through QPCs within the local Ohm's law. In this work, we compute the potential and current distributions of serially connected QPCs, starting from the calculation of bare electrostatic potential assuming a crystal structure which is used experimentally. Next, applying a perpendicular magnetic field to the 2DES, we obtain the spatial distribution of current-carrying channels, depending on field strength. In the final investigation, an external in-plane electric field is taken into account, and the current flow is obtained under certain conditions. The results of this study indicate that minor field variations are robust in determining the current distribution, whereas gate potential $V_{\rm G}$ and temperature $T$ dominate the scattering processes, as expected.

\section{Model}

\begin{figure}
	\centering
	\includegraphics[width=5cm]{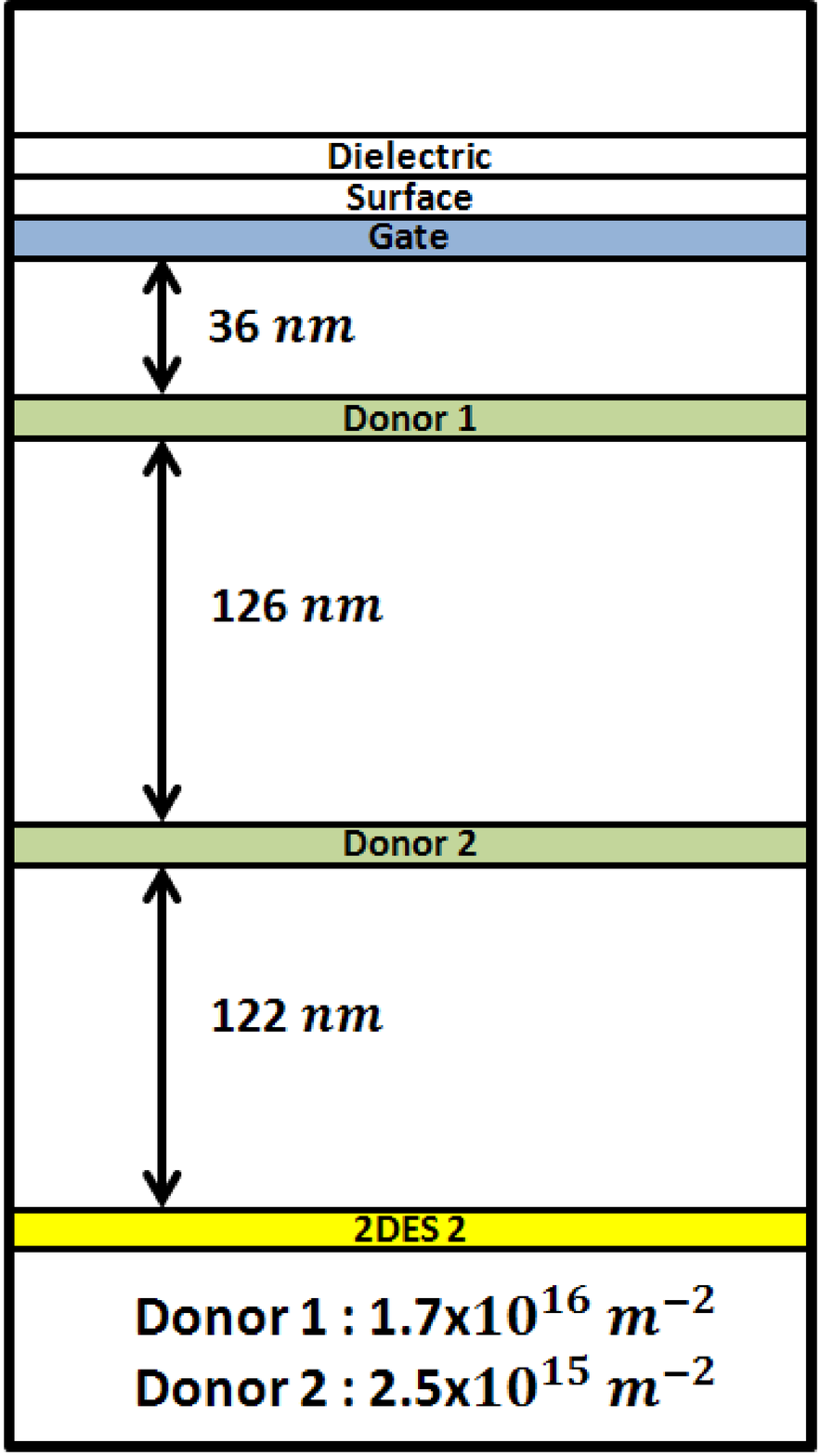}
	\caption{\label{fig1} Schematic presentation of the GaAs/AlGaAs heterostructure. The crystal is in vacuum and 2DES is formed at the interface of the junction. The number of donors and the structure geometry are taken from experimental reports~\cite{2013NatSR,2017NatCo}.}
\end{figure}

Since the main goal of our study is to obtain the current distribution through the QPCs, one should first obtain the electrostatic potential distribution $V(x,y,z)$ (hence , electron density distribution, $\rho(x,y,z)$) via solving the Poisson equation,

\begin{equation}\label{key}
 \nabla^{2}V(x,y,z)= - 4\pi\rho(x,y,z)
\end{equation}
 in 3D by imposing relevant boundary conditions and using  material properties. For this purpose, we utilize a well-developed numerical method called EST3D , which is based on an iterative method to obtain, self-consistently, $V(x,y,z)$ and $\rho(x,y,z)$~\cite{Arslan08:125423}. An advanced 3D fast-Fourier subroutine is used to calculate the distributions, layer by layer, where all the surfaces (top, side  and bottom) are assumed to be under vacuum, silicon-doped (two layers of delta-doping) GaAs/AlGaAs heterostructure is considered (see Fig . \ref{fig1}) and metallic gates are defined on the surface, which are kept at $V_{\rm G}$, as in our previous studies~\cite{Salman:13,Atci:17}. The vacuum, the heterostructure and the metallic gates are defined by their dielectric constants. Initially, the delta-doped silicon layers are charged positively with a fixed number of charges depending on the crystal growth parameters. The metallic gates are taken to be charged positively or negatively. The rest of the heterostructure, i.e., surfaces including  vacuum, are neutral in charge. Starting with these boundary conditions, one obtains $V(x,y,z)$ and $\rho(x,y,z)$ depending on the potential on gates, strength of doping and thickness of GaAs and AlGaAs layers.

Equipped with the self-consistently calculated potential and charge density distributions for each layer at zero temperature and magnetic  field, one can obtain the density and current distributions in the presence of external in-plane electric and off-plane magnetic fields, using the Newton-Raphson iteration~\cite{Eksi:10,Yildiz:14,Kilicoglu16:035702}. Our strategy is to use $V(x,y,z)$ as an initial input, obtained in the previous step, and calculate finite temperature and magnetic field reconstructed potential and charge distributions. Considering the experimental values of energies and charge densities, it can be easily seen that our results are viable, such that the typical charge density of the 2DES is similar to $3\times10^{15}$ m$^{-2}$ , corresponding to a Fermi energy ($E_F$) of 13 meV. At $10$ Tesla magnetic energy, $\hbar \omega_{c}(~ \omega_{c}=eB/m^{*}$) is   on the order of 17 meV, and thermal energy (T $\leq$ 10 K) is much smaller than the confinement energy (approximately 4 eV) and potential (energy) on metallic gates ($\sim$ -0.2 eV). The details of the calculation procedures and validity of the assumptions are explained in our previous studies ~\cite{Arslan08:125423,Kilicoglu16:035702}.

While performing calculations considering an external current, we always stay in the linear-response regime, which essentially imposes that the applied in-plane electric field does not affect the density and potential distributions. This is well justified, as the current amplitudes considered are much smaller than the Fermi energy~\cite{Guven03:115327}. In the following Sections we present our numerical results, first investigating a toy model using cosine-defined QPCs. The rationale is to clarify the effect of scattering processes without including the geometrical dependencies on them and explain the current distribution depending only on $B$ field. Next, we calculate the same quantities for higher gate voltage QPCs at various magnetic fields.

\begin{figure}
	\centering
	\includegraphics[width=7.5cm]{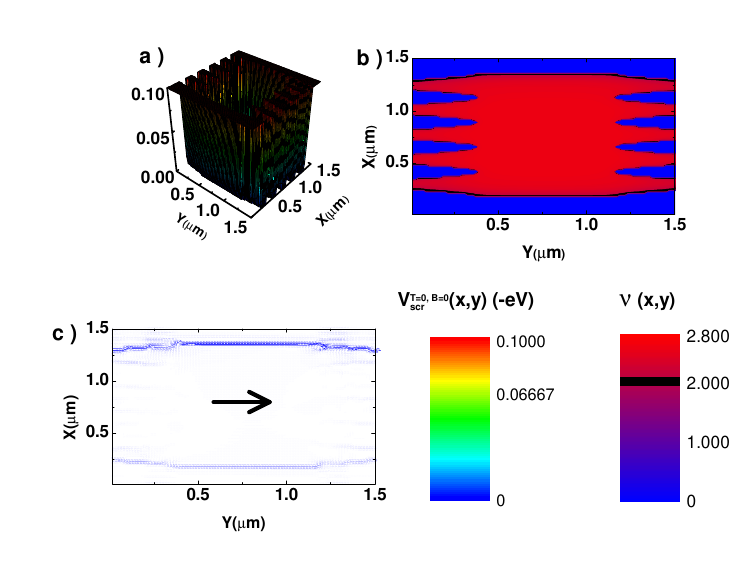}
	\caption{\label{fig2} (a) The spatial distribution of screened potential at zero temperature and vanishing magnetic field with $V_{g}=-0.1$  V (b) Self-consistent filling factor distribution and (c) The current distribution at $B= 7.5$ T. Gray scale  on the left legend denotes the potential strength, whereas right scale shows the filling factor with black indicating $\nu=2$. Small light blue arrows  depict local current distribution and large black arrow shows total current direction. In (b) and (c), equilibrium temperature is $7.43$ K.
}
\end{figure}

\begin{figure}
	\centering
	\includegraphics[width=7cm]{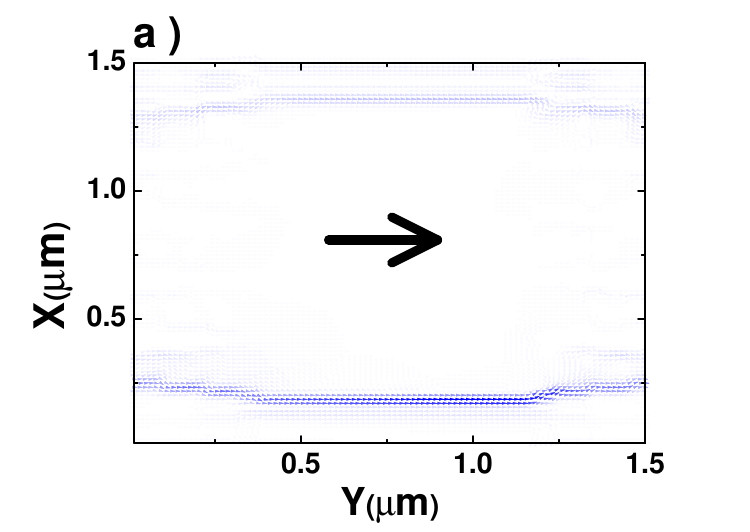}
	\includegraphics[width=7cm]{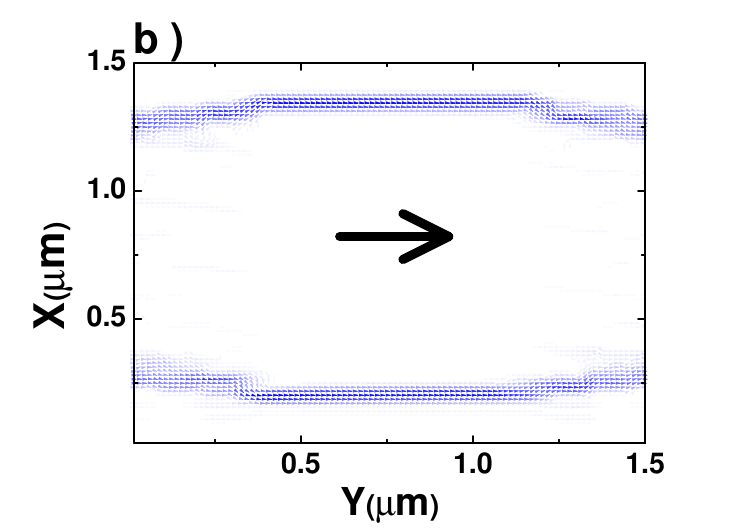}
	\caption{\label{fig3} Current distributions at (a) $8$ T and (b) $8.5$ T, resulting in $T_{\rm E}$ $10.15$ K and $2.0$ K, respectively. At the higher $B$ value, most of the current is owing  without scattering; hence, $T_{\rm E}$ is reduced.
}
\end{figure}

\section{Results and discussion}

It is significant to compare our numerical results with already existing ones , to show the consistency between them. The usual approach is to assume that the QPCs generate cosine- or Gaussian-like potentials  in the plane of the 2DESs~\cite{Macucci02:39,Igor07:qpc1,Igor07:qpc2}. By such modeling, one can obtain reliable results without further computational complications compared to realistically modeled devices. Here, we prefer to use cosine functions because fast-Fourier transformation processes are much faster and more precise compared to other well-defined functions. Also note that we are only interested in the transport properties of the 2DESs; hence, we show our numerical results just for the $z=z_{\rm 2DES}$ layer, i.e., 284 nm below the gate. Therefore, when the electron (number) density $n_{\rm el}(x,y)$ is presented, we take the result of 3D calculation for $\rho(x,y,z_{\rm 2DES})$ . A similar path is taken for the screened potential at zero temperature and vanishing $B$ field, namely $V^{T=0,B=0}_{\rm scr}(x,y,z_{\rm 2DES})$.

As an illuminating example, we define four QPCs on the top surface of our heterostructure, as shown in Fig. \ref{fig2}. The corresponding screened potential profile ($V^{T=0,B=0}_{\rm scr})$) is shown in Fig. \ref{fig2}a, together with the dimensionless electron density ($\nu(x,y)$, Fig. \ref{fig2}b) and current ($j(x,y)$) distribution as a function of position in the plane of the 2DES (i.e. $z=z_{2DES}$), Fig . \ref{fig2}c. It is beneficial to parametrize density by normalizing it with the strength of the external magnetic field. The dimensionless electron density is called the filling factor and is given by $\nu(x,y)=2\pi \ell^{2}n_{\rm el}(x,y)$, where $\ell$ is the magnetic length, defined as $\ell^{2}=eB/h$. From  Fig. \ref{fig2}a, one can see that the potential generated by the surface gates (both QPCs and side gates, dark blue regions in Fig. \ref{fig2}b) depletes electrons beneath them ($V_{scr} (x,y)$= -0.1 eV), and the external potential is well screened by the electrons elsewhere ($V_{scr} (x,y)\simeq 0.0$ eV), if a repulsive potential is applied to the gates ($V_{\rm G}=-0.1$ eV). Obviously, the QPCs constrain electron transport together with the side gates which confine them to a quasi-2D channel. The resulting density distribution is shown in Fig. \ref{fig2}b, where the color gradient presents the variation, and regions without a gradient (dark blue) indicate the electron depleted zones below the gates.

It is significant to emphasize that integer filling factors play a distinguishing role both in screening and transport properties of the system at hand. Let us consider a situation where the ratio between self-consistent electron density and the magnetic flux density assumes an integer value. In this case, the Fermi energy falls in between the magnetic field quantized density of states (DOS) locally; hence, there are no states available at these regions. This leads to areas of poor screening and constant electron density, called the incompressible strips~\cite{Chklovskii92:4026}. On the other hand, for the very same reason, scattering is suppressed, leading to a highly reduced resistance along the current direction. Essentially, local longitudinal resistance vanishes at the limit of zero temperature ~\cite{,Guven03:115327,siddiki2004}. Therefore, we depicted integer filling factors by black color, $\nu=2$, to  estimate locations of the incompressible strips.

\begin{figure*}[t!]
	\centering
	\includegraphics[width=5cm]{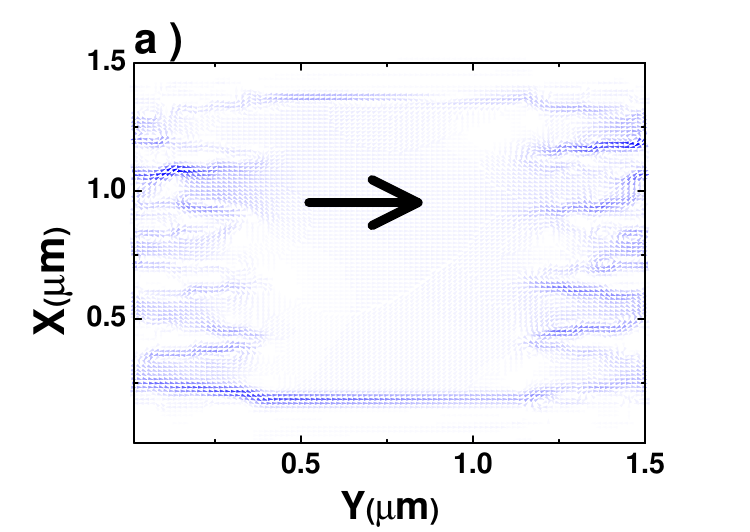}
	\includegraphics[width=5cm]{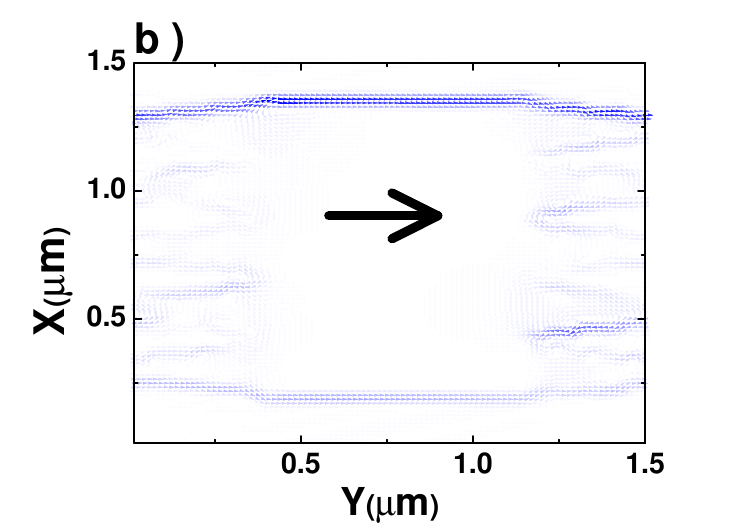}
	\includegraphics[width=5cm]{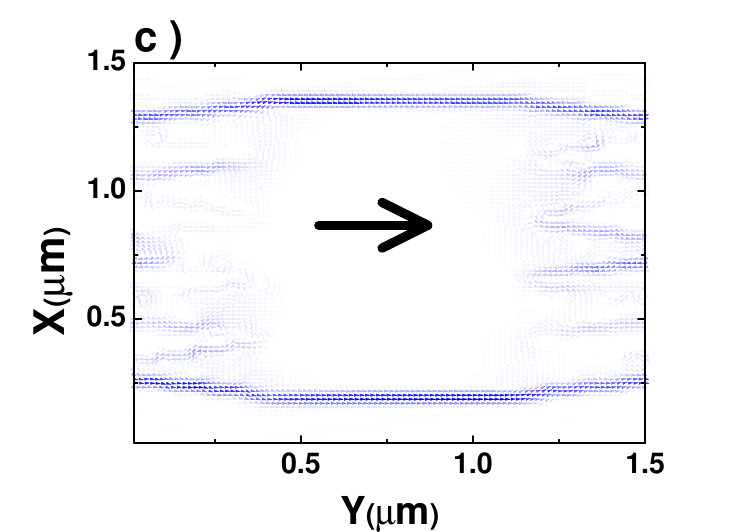}
	\includegraphics[width=5cm]{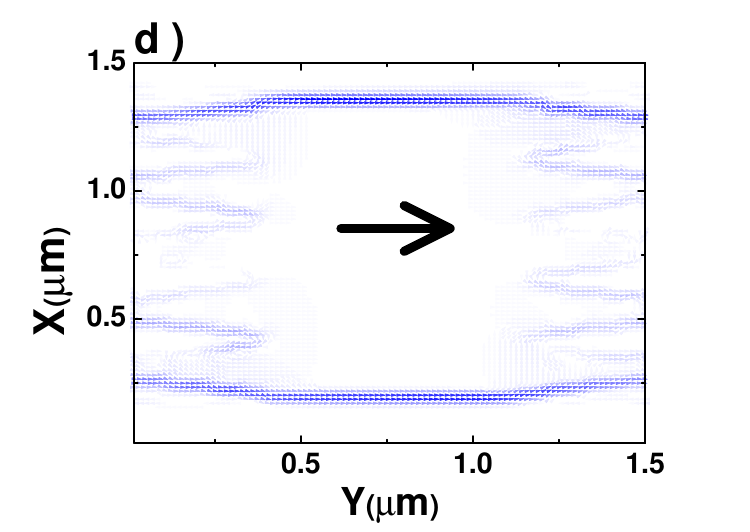}
	\includegraphics[width=5cm]{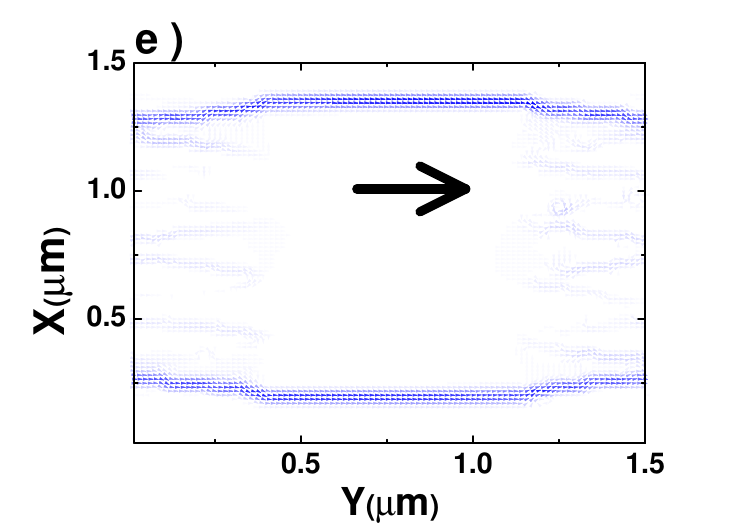}
	\includegraphics[width=5cm]{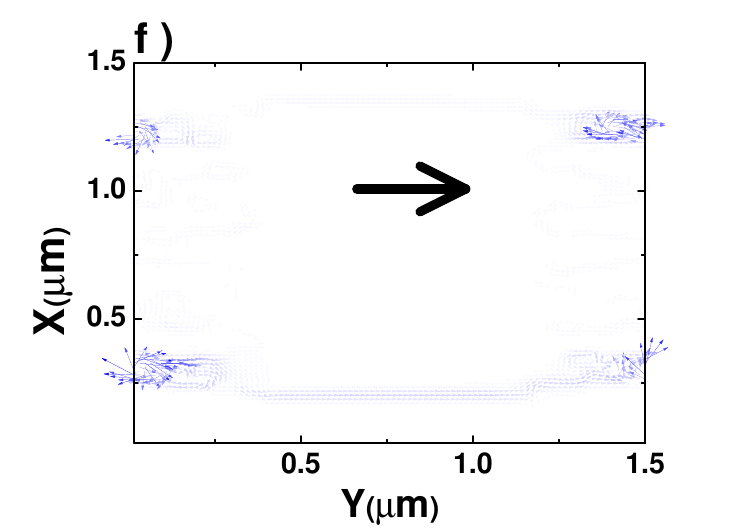}
	\caption{\label{fig4} Current distributions  where $V_{g}=-0.2$ V at various $B$ values and equilibrium temperatures: (a) $B=6.0$ T, $T_{\rm E}= 4.55$ K, (b) $B=6.5$ T, $T_{\rm E}= 3.05$ K, (c) $B=7.0$ T, $T_{\rm E}= 3.66$ K, (d) $B=7.5$ T, $T_{\rm E}= 5.12$ K, (e) $B=8$ T, $T_{\rm E}= 3.48$ K and (f) $B=8.5$ T, $T_{\rm E}= 4.2$ K. The strongly nonlinear change of equilibrium temperature is a direct consequence of Thomas-Fermi screening~\cite{Oh97:108}.
}
\end{figure*}

Fig. \ref{fig2}c shows the distribution of current that is imposed in the positive $y$ direction, with a normalized amplitude of $0.01$. We observe that most of the current exerted  is confined to integer-filling-factor regions, namely, $\nu$=2. A closer look at the data indicates that some of the current is backscattered in the proximity of the top- and bottom-most QPCs, indicating that because of finite temperature, a negligible number of transport electrons ($<0.01 \% $) are scattered to compressible (with high DOS, metal-like)  regions, where longitudinal resistance is finite. A remarkable feature in the current distribution is the asymmetry between the upper and lower parts of Fig. \ref{fig2}c. It is seen that more current flows from  the upper half. This is in agreement with the experimental~\cite{afif:njp2} and theoretical~\cite{SiddikiEPL:09} findings reported in the literature, justifying our results. The main reason for such behavior is grounded in the symmetry-breaking external magnetic field, namely, the Lorentz force resulting in induced Hall voltage. Keep in mind that, at relatively low magnetic field , the edge incompressible strips  are as narrow as the Fermi wavelength; hence, at this magnetic field the current is mostly driven by the drift velocity. Therefore, one can conclude that scattering is mainly dominated by impurities. At elevated field strengths (Fig. \ref{fig3}a and Fig. \ref{fig3}b), we observe that current first shifts to the lower part of the sample ($B=8$ T) and then is approximately symmetrically distributed over the sample at $B=8.5$ T. This is mainly due to the enlargement of the incompressible strips while increasing the magnetic field. At $B=8.0$ T, the lower incompressible strip is well developed; i.e., the width of the strips is larger than the Fermi wavelength. Hence, the current is confined mainly to this scattering-free region. Once the magnetic field strength is increased by $0.5$ T, both incompressible strips at the lower and higher parts of the sample become larger than the Fermi wavelength; therefore, current is shared between them in an approximately equal manner. We are able to confirm this behavior just by checking the convergence temperature of the system. It is observed that, for antisymmetric current distributions, the dissipation is higher; hence, equilibrium (convergence) temperature is $7.43$ K and $10.15$ K for $7.5$ T and $8.0$ T, respectively . In accordance with our conclusion, the equilibrium temperature  $T_{\rm E}$ is lower once almost all of the current is confined to well-developed incompressible strips, namely, $2.0$ K at $8.5$ T.

Before presenting further results, to summarize the main ow of our understanding: The current is confined to the scattering-free incompressible strips, where dissipation is suppressed, leading to lower equilibrium temperatures. Depending on magnetic field, the existence, location and widths of the strips vary such that at lower fields the upper strip, at intermediate fields the lower strip and at higher fields both strips are well developed.

Next, we present results where the depleting gates are biased with a higher negative voltage of $-0.2$ V, yielding a steeper screened potential profile, which in turn leads to narrower incompressible strips. Fig. \ref{fig4}a-f presents our numerical results considering six characteristic $B$ values, in increasing order. At the lowest field value ($6$ T), no incompressible regions are formed; therefore, current is distributed all over the sample with high dissipation yielding a high $T_{\rm E}$ $(= 4.55$ K). Similar to the previous situation, the first incompressible strip is formed at $6.5$ T at the upper edge of the sample, where relatively more current is confined to the strip. This yields less scattering, which ends up with the lowest $T_{\rm E}$. Current is approximately symmetric at $7.0$ T, but interestingly, $T_{\rm E}$ is higher compared to the previous case. One can interpret this behavior as follows: although the current is confined to incompressible strips at both sides of the sample, the remaining current is highly scattered between the QPCs, increasing the dissipation, which then elevates the temperature. Our interpretation is justified when we consider two consecutive field strengths, namely, $7.5$ T and $8.0$ T. In both cases, the current distribution is symmetric; however, at $7.5$ $T_{\rm E}$ the temperature is $5.12$ K, being the highest value of our interval of interest, and then decreases to $3.48$ K at $8.0$ T, where bulk scattering is suppressed, as can be seen clearly from Fig. \ref{fig4}e. At the highest $B$ strength, although the bulk scattering is predominantly suppressed, the strong back-scattering at the injection (bottom left of the sample) and the collection (top right) regions dissipation is high , and as a consequence, $T_{\rm E}$ increases to $4.2$ K.

The results shown above give sufficient  information about the current distribution in the close vicinity of the QPCs depending on the magnetic field. Moreover, taking into account the scattering processes, one can also comment  on the variation of the equilibrium temperature, depending both on $B$ field strength and also on the formation of incompressible strips , strongly bound to the steepness of the potential profile determined by the gate voltage $V_{\rm G}$.

\section{Conclusion}
Our investigation focuses on the current distribution of parallel-configured QPCs under QH conditions, utilizing self-consistent numerical calculation schemes. We obtain the potential and charge distribution of a GaAs/AlGaAs heterojunction by solving the Poisson equation in 3D for given boundary conditions. The obtained potential at the layer of 2DES is used to calculate the distribution of scattering-free incompressible strips and hence the current. We found that the location of the transporting channels depends strongly on the applied perpendicular magnetic field strength and switch from asymmetric to symmetric. This behavior is also observed once the gate voltages are varied.

Our main finding is that the variation of the equilibrium temperature is due to (both back- and forward-) scattering affecting the dissipation. In conclusion, low equilibrium temperatures are obtained if the current is mainly confined to the scattering-free incompressible strips.

Our findings are in accordance with previous theoretical and experimental studies. Moreover, we are able to demonstrate that, in a parallel configuration of cosine-defined QPCs, equilibrium temperature is a key indicator to determine coherent transport through such quantum devices. It is admirable to obtain similar results for realistically determined QPCs and to compare them with experimental results. It is also a challenging problem to include (Joule) heating effects, in order to investigate the dissipation processes microscopically.

\section*{Acknowledgement} A.S. thanks Mimar Sinan Fine Arts University Physics Department members for fruitful discussions on theoretical issues.

\end{document}